\def   \ni {\noindent}

\def   \ssk {\vskip  5truept}

\def   \bsk {\vskip 15truept}

\def   \newpage {\vfill\eject}
\def   \newline {\hfil\break}

\documentstyle[epsfig]{article}
\begin{document}

\hsize 5truein
\vsize 8truein
\font\abstract=cmr8
\font\keywords=cmr8
\font\caption=cmr8
\font\references=cmr8
\font\text=cmr10
\font\affiliation=cmssi10
\font\author=cmss10
\font\mc=cmss8
\font\title=cmssbx10 scaled\magstep2
\font\alcit=cmti7 scaled\magstephalf
\font\alcin=cmr6
\font\ita=cmti8
\font\mma=cmr8
\def\ref{\par\noindent\hangindent 15pt}
\null


\title{\ni INTEGRAL: THE CURRENT STATUS}

\bsk \bsk
\author{\ni Christoph~Winkler}

\bsk
\affiliation{\ni Space Science Department of ESA, Astrophysics Division,
ESTEC, 2200AG Noordwijk, The Netherlands}

\bsk
\baselineskip = 12pt

\abstract{ABSTRACT \ni The International Gamma-Ray Astrophysics
Laboratory
(INTEGRAL) is dedicated to the fine spectroscopy
($\Delta$E: 2 keV FWHM @ 1.3 MeV) and
fine imaging (angular resolution: 12$^{\prime}$ FWHM)
of celestial gamma-ray sources in
the energy range 15 keV to 10 MeV with concurrent
source monitoring in the X-ray (3 - 35 keV) and optical (V, 550 nm) range. 
The mission is conceived as an observatory led by ESA  with contributions from
Russia and NASA.
The INTEGRAL observatory will provide to the science community at large 
an unprecedented combination of imaging and spectroscopy over a wide 
range of energies.
Most of the observing time will be open to the scientific community. 
This paper summarises the key scientific goals of the mission, the
current development status of the payload and
spacecraft and it will give an overview of the science ground 
segment including data centre, science operations and the 
key elements of the observing programme.}

\bsk
\baselineskip = 12pt

\bsk
\baselineskip = 12pt


\text{\ni 1. INTRODUCTION
\ssk
\ni

The International Gamma-Ray Astrophysics
Laboratory
(INTEGRAL) is dedicated to the fine spectroscopy
($\Delta$E: 2 keV FWHM @ 1.3 MeV) and
fine imaging (angular resolution: 12$^{\prime}$ FWHM)
of celestial gamma-ray sources in
the energy range 15 keV to 10 MeV.
The INTEGRAL observatory
will provide to the science community at large an unprecedented combination
of imaging and spectroscopy over a wide range of X-ray and gamma-ray
energies including optical monitoring.
The mission is conceived as an observatory led by
ESA with contributions from
Russia and NASA and will be launched in 2001.
ESA is responsible for the overall spacecraft and mission design,
instrument integration into the payload module,
spacecraft integrations and testing, spacecraft operations including one
ground station, science operations,
and distribution of scientific data.
Russia will provide a PROTON launcher and launch facilities, 
and NASA will provide ground station
support through the Deep Space Network.
The scientific instruments and the INTEGRAL Science Data Centre will be 
provided by large collaborations from many
scientific institutes in almost all ESA member states, USA, 
Russia, Czech Republic 
and Poland, nationally funded,
and led by Principal Investigators (PI's).
\newpage

\bsk
\ni 2. SCIENTIFIC OBJECTIVES
\ssk
\ni

INTEGRAL is a 15 keV - 10 MeV gamma-ray mission with concurrent source
monitoring at X-rays (3 - 35 keV) and in the optical range (V, 500 - 600 nm).
All instruments  - co-aligned with large FOV's - cover simultaneously
a very broad energy range of high energy sources (Tables 1, 2).

The scientific goals of INTEGRAL will be attained by fine spectroscopy 
with fine
imaging and accurate positioning of celestial sources of gamma-ray emission.
Fine spectroscopy over the entire energy range will permit spectral features
to be uniquely identified and line profiles to be determined for physical
studies of the source region. The fine imaging capability of INTEGRAL within
a large field of view will permit the accurate location and hence
identification of the gamma-ray emitting objects with counterparts at other
wavelengths, enable extended regions to be distinguished from point sources
and provide considerable serendipitous science which is very important for
an observatory-class mission.
In summary the scientific topics will address:
(i) compact objects, (ii) stellar nucleosynthesis,
(iii) high energy transients, 
(iv) mapping of diffuse continuum and line emission, 
(v) the galactic Centre,
(vi) particle processes and acceleration,
(vii) transrelativistic pair plasmas
(viii) nearby galaxies, clusters of galaxies, AGN, cosmic diffuse background,
(ix) identification of high energy sources,
(x) compilation of unidentified gamma-ray objects as a class,
PLUS: (xi) unexpected discoveries.

In particular our present knowledge of the 
galactic Centre region, a classical target for
high energy astrophysics, is at these energies largely based on the 
results on (mostly variable) point sources detected in hard X-rays/low energy 
gamma-rays by the GRANAT instruments SIGMA
(Vargas et al. 1997) and ART - P (Pavlinsky et al. 1994),
as well as the results on diffuse galactic emission as measured by 
the CGRO instruments COMPTEL 
at 1.8 MeV of Al$^{26}$ (Diehl et al. 1995) and OSSE at 511 keV
(Purcell et al. 1997).
The following (incomplete) 
list shows what can be studied in great detail with 
INTEGRAL:\\
\ni$\bullet$ map diffuse 511 keV and 1.8 MeV emission on large angular
scale\\ 
$\bullet$ measure the spectrum of the diffuse Galactic continuum emission\\ 
$\bullet$ image known sources, identify new point sources, companions of 
compact sources and hot spots from diffuse mapping in the G.C.
region with high accuracy\\
$\bullet$ determine the point source contribution to the observed 511 keV flux\\
$\bullet$ study the continuum characteristics of an ensemble of point sources
(spectral shape of neutron stars vs black hole candidates)\\
$\bullet$ perform spectroscopic studies of individual point sources: 
narrow lines (cyclotron and nuclear lines), broad lines 
(511 keV features including 
backscattering lines), line shifts (gravitational redshift ?), line shapes
and line profiles \\
$\bullet$ perform continuum and line studies of ''hot spots`` 
identified in diffuse maps, i.e. the cosmic ray/dust ratio and 
cosmic ray/gas ratio in case of narrow and broad lines, respectively, 
with line profile analysis and isotope determination \\
$\bullet$ timing, variability (QPO's) and 
polarisation analysis of compact sources\\

\newpage
\ni 3. SCIENTIFIC PAYLOAD
\ssk
\ni

\begin{table}[htp]
\footnotesize
\begin{center}
\begin{tabular}{|l|l|l|}
\hline
Instrument& Energy range & Main purpose\\ \hline
Spectrometer SPI & 20 keV - 8 MeV &  Fine spectroscopy of narrow lines \\
 & &  Study diffuse emission on $>$deg scale \\ \hline
Imager IBIS & 15 keV - 10 MeV &  Accurate point source imaging\\
& &  Broad line spectroscopy and continuum \\ \hline
X-ray Monitor JEM-X & 3 - 35 keV &  Source identification\\
& &  Monitoring @ X-rays \\ \hline
Optical Monitor OMC & 500 - 600 nm &  Optical monitoring of \\
& & high energy sources \\ \hline
\end{tabular}
\end{center}
\caption{Table 1: INTEGRAL science and payload complementarity}
\end{table}
\normalsize

The INTEGRAL payload consists of two main gamma-ray instruments: Spectrometer
SPI and Imager IBIS, and of two monitor instruments, the X-ray Monitor
JEM-X and the Optical Monitoring Camera OMC.
The design of the INTEGRAL instruments is largely driven by the 
scientific requirement to establish a payload of scientific complementarity.
As shown in Table 1, the payload does meet this goal.

Each of the main gamma-ray instruments, SPI and IBIS, 
has both spectral and angular resolution, but they are
differently optimised in order to complement each other and to achieve overall
excellent performance.
The two monitor instruments (JEM-X and OMC) will provide complementary
observations of high energy sources at X-ray and optical energy bands.
An overview of the INTEGRAL payload is given below, detailed
descriptions can be
found in the various instrument papers presented at this workshop.
Also part of the payload is a small particle radiation monitor, which
continuously measures the particle environment of the spacecraft. 
Therefore it is possible to provide essential
information to the payload in case high particle background (radiation belts,
solar flares) is being encountered. This information is used to
decide on switch - off and switch - on of instrument high voltages and
to provide actual background information for sensitivity estimates.

\noindent
{\em Spectrometer SPI}\\
The Spectrometer SPI (Table 2) will perform spectral
analysis of gamma-ray point sources and
extended regions with an unprecedented
energy resolution of 2 keV (FWHM) at 1.3 MeV.
This will be accomplished using an array of 19 hexagonal high purity
Germanium detectors cooled by two pairs of Stirling Coolers to
85 K. The total detection area is 500 cm$^2$.
A hexagonal coded aperture mask is located 1.7 m
above the detection plane in order to image large regions of the sky
(fully coded field of view = 16$^{\circ}$) with an angular resolution of
2$^{\circ}$. 
In order to reduce background radiation, the detector assembly
is shielded by an active BGO  veto system which extends around the
bottom and side of the detector almost completely up to the coded mask.
A plastic veto between mask and upper veto shield ring further reduces
background events. 

\noindent
{\em Imager IBIS}\\
The Imager IBIS (Table 2)
provides powerful diagnostic capabilities of fine imaging (12$^{\prime}$ 
FWHM), source identification and spectral sensitivity to both continuum and
broad lines over a broad (15 keV - 10 MeV)
energy range. The energy resolution is  7 keV @ 0.1 MeV and 60 keV @ 1 MeV.
A tungsten coded aperture mask (located at 3.2 m above
the detection plane)
is optimised for high angular resolution imaging.
Sources ($> 10\sigma$) can be located to $< 60''$.
As diffraction is negligible at gamma-ray wavelengths, the angular resolution
obtainable with a coded mask telescope is limited by the spatial resolution
of the detector array. The IBIS design takes advantage of this by utilising
a detector with a large number of spatially resolved pixels, implemented
as physically distinct elements.
The detector uses two planes, a front layer (2600 cm$^2$)
of CdTe pixels, each (4x4x2) mm, and a second one (3100 cm$^2$)
of CsI pixels, each (9x9x30) mm. 
The division into two layers allows the paths of the photons
to be tracked in 3D, as they scatter and interact with more than one element.
The aperture is restricted
by a thin passive shield. The detector array
is shielded from the sides and below by an
active BGO veto.

\noindent
{\em X-Ray Monitor JEM-X}\\
The Joint European X-Ray Monitor JEM-X (Table 2)
supplements the main INTEGRAL instruments (Spectrometer SPI and Imager
IBIS) and
plays a crucial role in the detection and identification of the gamma-ray
sources and in the analysis and scientific interpretation of INTEGRAL
gamma-ray data.
JEM-X will make observations simultaneously with the main
gamma-ray instruments and provides images with 3$^{\prime}$ angular
resolution in the 3 - 35 keV prime energy band.

\begin{table}[htp]
\footnotesize
\begin{flushleft}
\begin{tabular}{|l|l|l|l|l|}
\hline
& SPI & IBIS & JEM-X & OMC \\ \hline
Energy range & 20 keV - 8 MeV & 15 keV - 10 MeV & 3 keV - 35  keV 
& (500 - 600) nm\\ \hline
Detector area (cm$^2$) & 500  & 2600  (CdTe) & 1000 (2 units & CCD\\
& & 3100 (CsI) &  each 500 cm$^2$)& (2048$\times$1024 pxl)\\ \hline
Spectral resolution & 2 (at 1.3 MeV)&  7 (at 100 keV) 
& 1.5  @ 10 keV   & -- \\
FWHM (keV) & & 60 (@ 1 MeV) &  & \\ \hline
Field of view  & 16$^{\circ}$  & 9$^{\circ}$ x 9$^{\circ}$  &
4.8$^{\circ}$  & 5.0$^{\circ}$ x 5.0$^{\circ}$ \\ \hline
(fully coded) & & & & \\ \hline
Angular resolution & 2$^{\circ}$ FWHM & 12$^{\prime}$ FWHM &
3$^{\prime}$ FWHM & 17.6$^{\prime}$$^{\prime}$/pixel \\ \hline
Typical source location & 20$^{\prime}$ & $<$ 1$^{\prime}$ 
&$<$ 20$^{\prime}$$^{\prime}$ & $<$ 8$^{\prime}$$^{\prime}$\\ \hline
Continuum sensitivity$^1$ & 7$\times$10$^{-8}$ &
4$\times$10$^{-7}$ &
1$\times$10$^{-5}$ &
19.7$^{m_v}$ \\
(3$\sigma$, 10$^6$ s) &  @ 1 MeV &  @ 100 keV &  @ 6 keV & 
(3$\sigma$, 10$^3$ s)\\ \hline
Line sensitivity$^1$ & 5$\times$10$^{-6}$ &
1$\times$10$^{-5}$  &
2$\times$10$^{-5}$  &
--\\
(3$\sigma$, 10$^6$ s) &  @ 1 MeV &
@ 100 keV & 
@ 6 keV & \\ \hline
Timing accuracy (3$\sigma$) & 100 $\mu$s & 67$\mu$s -- 1000 s  &
128 $\mu$s& $>$ 1s \\
Mass (kg) & 1309 &628 &65 & 17\\ \hline
Power (W)  & 373  & 275  &55 & 18\\ \hline
Data rate (kbps)  & 20 (avge) & 57 (avge) &7 &2\\
\hline
\end{tabular}
\caption{Table 2: Key parameters of the INTEGRAL scientific payload. (1) Units
of sensitivities are (ph cm$^{-2}$ s$^{-1}$ keV$^{-1}$) for continuum and
(ph cm$^{-2}$ s$^{-1}$) for lines.}
\end{flushleft}
\end{table}
\normalsize

The photon
detection system consists of two identical 
high pressure imaging
microstrip gas chambers (Xenon at 5 bar) each viewing the sky through
a coded aperture mask (4.8$^{\circ}$ fully coded FOV), 
located at a distance of 3.2 m above the detection
plane. The total detection area is 1000 cm$^2$.

\noindent
{\em Optical Monitoring Camera OMC}\\
The Optical Monitoring Camera OMC (Table 2) consists of a 
passively cooled CCD
in the focal plane of a 50 mm lens. The CCD (1024 x 2048 pixels) uses
one section (1024 x 1024 pixels) for imaging, the other one for frame transfer
before readout. The FOV is 5$^{\circ}$ $\times$ 5$^{\circ}$ with a pixel size
of $17.6''$.
The OMC will observe the optical emission from the prime targets of the
INTEGRAL main gamma-ray instruments with the support of the X-Ray Monitor
JEM-X.  Variability
patterns on timescales of 1 s and longer, up to months and years will be 
monitored. The limiting magnitude of 19.7$^{m_v}$ (3$\sigma$, 10$^3$ s), 
corresponds to $\sim$40 photons cm$^{-2}$s$^{-1}$keV$^{-1}$ (@ 2.2 eV) 
in the V-band.
Multi-wavelength
observations are particularly important in high-energy astrophysics where
variability is typically rapid. The wide band observing opportunity
offered by INTEGRAL is of unique importance in providing for the first
time simultaneous observations over seven orders of magnitude in
photon energy for some of the most energetic objects in the Universe.

\bsk
\ni 4. MISSION SCENARIO
\ssk
\ni

\begin{figure}
\centerline
{\psfig{file=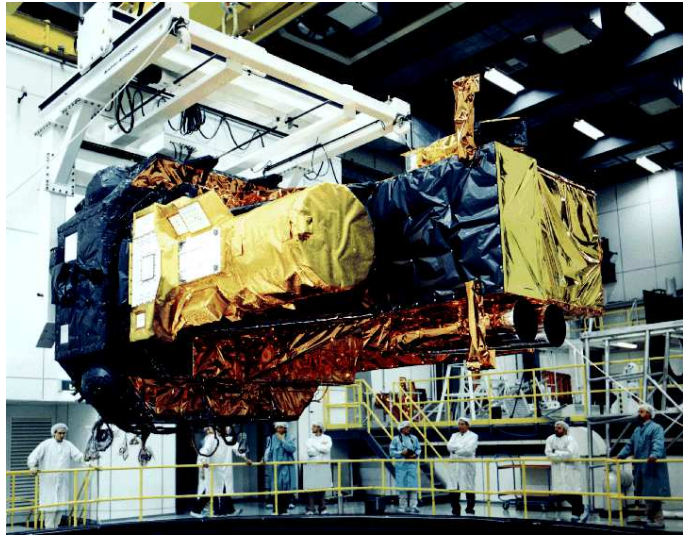, width=9cm}}
\caption{FIGURE 1. The structural \& thermal model of INTEGRAL 
during testing in ESTEC (Summer 98).
The cylindrically shaped SPI is next to the larger rectangular
payload module (PLM) structure housing the IBIS and JEM-X detectors inside. The
rectangular (golden) side of the PLM covers the IBIS and JEM-X coded masks.
Below the PLM are two star trackers, above the OMC. The black octagonal
structure to the left is the service module. Solar arrays are not shown.
}
\end{figure}
    
The INTEGRAL spacecraft 
consists of a service module (commonly designed with the service module of
the ESA XMM mission) containing all
spacecraft subsystems, and a payload module containing the scientific
instruments. 
During summer 1998, the service module and the payload have 
succesfully completed the structural and thermal test (STM, Figure 1)
programme and the electrical test (EM) programme is well underway.
Further details on the current spacecraft design can be found in
Carli et al. 1999).

INTEGRAL (with a payload mass of 2019 kg and a total launch mass of 
$\sim$4000 kg) will be launched
in April 2001 into a highly eccentric
orbit with high
perigee in order to provide long periods of uninterrupted observation with
nearly constant background and away from trapped radiation.
The baseline is to launch INTEGRAL with a Russian PROTON. ESA and
the Russian Space Agency have signed an arrangement in November 1997
formally securing the launcher for the mission.
The parameters for the orbit (see Carli et al. (1999) for further details) are: 
period 72 hours, inclination 51.6$^{\circ}$, 
initial perigee height 10 000 km, initial apogee height 153 000 km.
This orbit is a modification from the previous one (e.g. Winkler 1997)
in order to simplify both launcher operations and ground coverage, and 
for scientific reasons,
to maximize the time the spacecraft is spending above $\sim$40~000 to 60~000 
km based on recent analysis (Vargas 1998) of the radiation background 
observed by SIGMA/GRANAT at those high altitudes:
the particle background radiation affects  the performance
of high-energy detectors, and
scientific observations will be carried out while the spacecraft is above an
altitude of nominally 40~000 km.
However, the particle background of the local spacecraft 
environment will be continuously measured
by the on-board radiation monitor: this device allows the optimisation of the
observing time before or after radiation belt passages and solar flare
events, and provides essential information about the actual background.
Data from the onboard radiation monitor will be routinely checked
to verify and possibly update the nominal altitude above which scientific
observations will be performed.
A nominal altitude of 40~000 km implies that 90\% of the time spent 
on the orbit provided by PROTON can be used for scientific observations
However, a number of in-orbit activities have an influence on the net amount
of orbit time (e.g. slews, eclipses, resctrictive spacecraft operations,
instrument calibrations)
such that the average observation efficency becomes 85\% per year.
The real-time scientific data rate (including instrument housekeeping) 
has been recently increased by $\sim$30\% to 
86 kbps, an increase basically driven by scientific timing requirements
of the IBIS instrument.

The spacecraft employs fixed solar arrays: this means, that the 
target pointing of the spacecraft (at any point in time)
will remain outside an avoidance
cone around the sun. This leads to a minimum angle between any 
celestial source
and the sun of 50$^{\circ}$ during the nominal mission life (2 years) 
outside eclipse seasons and
60$^{\circ}$ during extended mission life (year 3+). During eclipse 
seasons (few weeks per year) 60$^{\circ}$ will be applied.

Because of imaging deconvolution requirements by SPI, the spacecraft
will routinely, during nominal operations, perform a series of
off-source pointing manouevres, known as ''dithering``.
These dithering patterns consist of sets of different
pointings at sky positions around the
nominal target position (at the centre).
The dithering points are separated by 2$^{\circ}$. The exposure
time per point is 20 minutes.
Two dither patterns will be employed: a 7 point hexagone and a 5$\times$5 point
raster, both centred on the target position. 
If required by observers, dithering can be disabled.

\bsk
\ni 5. GROUND SEGMENT
\ssk
\ni

The ground segment consists of two major
elements, the Operations Ground Segment (OGS) and the Science Ground Segment
(SGS):
The OGS, consisting of the ESA and NASA ground stations and ESA's Mission 
Operations Centre (MOC) at
ESOC 
will implement the observation plan received from the INTEGRAL Science
Operations Centre (ISOC) 
within the spacecraft system
constraints into an operational command sequence (Schmidt et al. 1999). 
In addition, the OGS
will perform all classical spacecraft operations, 
real-time contacts with spacecraft and payload, maintenance tasks and
anomaly checks
(i.e. including payload critical health and safety). MOC will determine the
spacecraft attitude and orbit, and will provide raw science data to the SGS.

The SGS itself consists of two centres, the
INTEGRAL Science Operations Centre (ISOC) and the
INTEGRAL Science Data Centre (ISDC).
The ISOC (Barr et al. 1999), provided by ESA and located
at ESTEC, will issue the AO for observing time and handle the 
incoming proposals. 
Accepted observation proposals will then be
processed at ISOC into an
optimised observation plan which consists of a timeline of target pointings
plus the corresponding instrument configuration.
This observation plan will then be forwarded to MOC to be uplinked
to the spacecraft. Furthermore, the ISOC  will validate any changes made
to parameters describing the on-board instrument configuration
and it will keep a copy of the scientific
archive produced at the ISDC. Finally, 
the ESA Project Scientist at the ISOC will decide on the generation of
TOO alerts (Target of Opportunity) in order to 
update and reschedule the observing programme.

The ISDC (Courvoisier et al., 1999), located in Versoix, Switzerland,
will receive the complete raw science telemetry plus the relevant ancillary
spacecraft data from the OGS/MOC. Science data will be processed,
taking into account the instrument characteristics, and raw data will be
converted into physical units.
Using incoming science and housekeeping information, the ISDC will routinely
monitor the instrument science performance and conduct a quick-look
science analysis.
Most of the
Targets of Opportunity (TOO) showing up during the lifetime of INTEGRAL
will be detected at the ISDC during the routine scrutiny of the data
and will be reported to ISOC. 
Scientific data products obtained by standard analysis tools
will be distributed to the observer and archived for later use by the
science community. 

At the time of writing, the software development for the ground segment
is proceeding according to schedule. 
Many elements of the software architecture are completing
the architectural design phase or are commencing detailed design phase.
%
            

\newpage
\ni 6. OBSERVING PROGRAMME
\ssk
\ni

INTEGRAL will be an observatory-type mission with a nominal lifetime of
2 years, an
extension up to 5 years is technically possible.
Most of the observing time
(65\% during year 1, 70\% (year 2), 75\% (year 3+))
will be awarded to the scientific community at large as the General Programme.
Typical observations will last from 10's of minutes up to two weeks. Proposals,
following a standard AO process, will be selected on 
their scientific merit only by a single Time Allocation Committee.
These selected
observations are the base of the General Programme.
The first call for observation proposals is scheduled for release at the 
end of 1999.
In principle, observers will receive data from all co-aligned and simultaneously
operating instruments onboard INTEGRAL.
The remaining fraction
of the observing time (i.e. 35\% (year 1), 30\% (year 2),  25\% (year 3+))
will be reserved, as guaranteed time, for the INTEGRAL Science Working Team 
for its contributions  to the
programme. 
This fraction, the Core Programme, will be devoted to: (i) a 
Galactic Plane Survey, (ii) a deep exposure of the central radian of the 
Galaxy, and (iii) pointed observations of selected regions/targets 
including TOO follow up observations.
The current status of the Core Programme 
is described in detail by Winkler et al. (1999). 
The full details of the Core Programme will be made available
at the issue of the first AO.
In accordance with ESA's policy on data rights, all scientific data will 
be made available to the scientific community at large
one year after they have been released to the observer.
This guarantees the use of the scientific
data for different investigations beyond the aim of a single proposal.

\bsk
\baselineskip = 12pt
{\abstract \ni ACKNOWLEDGMENTS
\ni
This paper has been written on behalf of the INTEGRAL Science Working Team:
G. Vedrenne (SPI), V. Sch\"onfelder (SPI), P. Ubertini (IBIS), F. Lebrun (IBIS),
N. Lund (JEM-X), A. Gimenez (OMC), T. Courvoisier (ISDC), N. Gehrels, 
W. Hermsen, J. Paul, G. Palumbo, S. Grebenev (Mission Scientists), B. Teegarden
(NASA/USA), R. Sunyaev (IKI/Russia) and C. Winkler (ESA).
For up-to-date information on INTEGRAL please consult the
INTEGRAL pages on the WWW: 
http://astro.estec.esa.nl/SA-general/Projects/Integral/integral.html

}

\bsk
\baselineskip = 12pt


{\references \ni REFERENCES
\ssk
\ref Barr P., et al., 1999, these proceedings
\ref Carli R., et al., 1999, these proceedings
\ref Courvoisier T., et al., 1999, these proceedings
\ref Diehl R., et al., 1995, A\&A 298, 445
\ref Pavlinsky M., et al. 1994, ApJ 425, 110
\ref Purcell W., et al., 1997, ApJ 491, 725
\ref Schmidt M., et al., 1999, these proceedings
\ref Vargas M., et al., 1997, ESA SP-382, 129
\ref Vargas M., 1998, ISDC internal report
\ref Winkler C., 1997, ESA SP-382, 573
\ref Winkler C., et al., 1999, these proceedings
}

\end{document}